\documentclass{article}

\usepackage{PRIMEarxiv}
\usepackage[show]{chato-notes}
\usepackage[utf8]{inputenc} % allow utf-8 input
\usepackage[T1]{fontenc}    % use 8-bit T1 fonts
\usepackage{hyperref}       % hyperlinks
\usepackage{url}            % simple URL typesetting
\usepackage{booktabs}       % professional-quality tables
\usepackage{amsfonts}       % blackboard math symbols
\usepackage{nicefrac}       % compact symbols for 1/2, etc.
\usepackage{microtype}      % microtypography
\usepackage{lipsum}
\usepackage{fancyhdr}       % header
\usepackage{graphicx}       % graphics
\graphicspath{{media/}}     % organize your images and other figures under media/ folder

\title{Cross-Lingual Query-Based Summarization of Crisis-Related Social Media: An Abstractive Approach Using Transformers
}

\author{
  Fedor Vitiugin \\
  Universitat Pompeu Fabra \\
  Barcelona, Spain \\
  \texttt{fedor.vitiugin@upf.edu} \\
   \And
  Carlos Castillo \\
  ICREA and Universitat Pompeu Fabra \\
  Barcelona, Spain \\
  \texttt{chato@icrea.cat} \\
}

\usepackage{paralist}

\usepackage{acronym}

\acrodef{BERT}{Bidirectional Encoder Representations from Transformers}
\acrodef{BiLSTM}{Bidirectional Long Short-Term Memory}
\acrodef{ERCC}{Emergency Response Coordination Centre}
\acrodef{GLOVE}{Global Vectors for Word Representation}
\acrodef{LASER}{Language-Agnostic SEntence Representations}
\acrodef{LSTM}{Long Short Term Memory}
\acrodef{MLP}{Multi-Layer Perceptron}
\acrodef{NER}{Named Entity Recognition}
\acrodef{NLP}{Natural Language Processing}
\acrodef{POS}{Part Of Speech}
\acrodef{SVM}{Support-Vector Machines}

\newcommand{\ourmethodclass}{CLiQC-CM\xspace}
\newcommand{\ourmethod}{CLiQS-CM\xspace}
\newcommand{\ourmethoddiv}{CLiQS-D-CM\xspace}

\newcommand{\ourmethodexpanded}{Cross-LIngual Query-based Summarization of Crisis Messages}
\newcommand{\ourmethodclassexpanded}{Cross-LIngual Query-based Classification of Crisis Messages}

\newcommand{\spara}[1]{\smallskip\noindent\textbf{#1}}

\begin{document}
\maketitle

\begin{abstract}
Relevant and timely information collected from social media during crises can be an invaluable resource for emergency management.
However, extracting this information remains a challenging task, particularly when dealing with social media postings in multiple languages.
This work proposes a cross-lingual method for retrieving and summarizing crisis-relevant information from social media postings.
We describe a uniform way of expressing various information needs through structured queries and a way of creating summaries answering those information needs.
The method is based on multilingual transformers embeddings.
Queries are written in one of the languages supported by the embeddings, and the extracted sentences can be in any of the other languages supported.
Abstractive summaries are created by transformers.
The evaluation, done by crowdsourcing evaluators and emergency management experts, and carried out on collections extracted from Twitter during five large-scale disasters spanning ten languages, shows the flexibility of our approach.
The generated summaries are regarded as more focused, structured, and coherent than existing state-of-the-art methods, and experts compare them favorably against summaries created by existing, state-of-the-art methods.
\end{abstract}

%%
%% The code below is generated by the tool at http://dl.acm.org/ccs.cfm.
%% Please copy and paste the code instead of the example below.
%%

%%
%% Keywords. The author(s) should pick words that accurately describe
%% the work being presented. Separate the keywords with commas.

%%
%% This command processes the author and affiliation and title
%% information and builds the first part of the formatted document.
\maketitle

\section{Introduction}

Social media platforms such as Twitter are widely used to share information during disasters and mass convergence events~\cite{castillo2016big}.
During these situations, users, including eyewitnesses, media, governmental and non-profit organizations, post an enormous volume of diverse content, from personal opinions and commentary to reports and messages providing relevant information that could lead to better situational awareness.
This work describes an approach to automatically summarize information posted in social media about an event, creating brief reports to help emergency response and recovery. %, such as casualties, damaged infrastructure, and other facts.
These reports can help emergency managers better understand a developing situation and plan the following actions accordingly~\cite{de2015geographic, huang2015geographic}.

The development of methods that automatically extract crisis-relevant information from social media has been an active line of work for many years~\cite{imran2015processing}.
Traditionally, crisis information extraction methods use linguistic and semantic resources mainly concentrated on one language \cite{rudra2015extracting}.
However, there are many cases where a single crisis affects several countries or regions that speak different languages~\cite{ventayen2017multilingual,8981479,lorini2019integrating}, or affects a region where the population speaks more than one language.

Previous work has shown that the information provided by social media postings is related to the language in which they are posted, and indeed messages in different languages about the same crisis often provide complementary information \cite{vitiugin2019comparison, lorinisocial}.
Extracting and summarizing information from social media in only one language introduces the risk of missing valuable information. 
However, creating or adapting language-specific resources or methodologies for new languages is expensive and time-consuming.
Therefore, current crisis informatics solutions need effective cross-lingual tools for extracting relevant information about appropriate categories of crisis-relevant information.

Our main contributions are:
\begin{itemize}
    \item We describe a flexible, query-based, cross-lingual method for collecting from social media relevant postings in multiple languages about specific information categories.
    The method uses pre-trained multilingual sentence embeddings (\acs{LASER} \cite{artetxe2019massively}) to extract postings from a general collection of crisis-related messages.
    \item We describe an approach for crisis summarization that takes as input relevant postings about an information category and generates a summary using a transformer-based language model (T5 \cite{raffel2020exploring}).
    We use clustering and diversification operations to create less redundant, more information-rich summaries.
\end{itemize}

We perform empirical validation using both crowdsourcing annotators and emergency management experts and release a new annotated dataset to evaluate multilingual crisis informatics systems.

The remainder of this paper includes a presentation of related work (\S\ref{sec:relatedwork}), followed by a description of the query-based method for crisis information extraction and cross-lingual classification and summarization (\S\ref{sec:method}). 
Next, we describe our experimental setup (\S\ref{sec:setup}) and the results of our analysis (\S\ref{sec:results}). 
Finally, we present our conclusions and envisioned future work (\S\ref{sec:conclusions}).

% - - - - - - - - - - Related Work - - - - - - - - - -

\section{Related Work}
\label{sec:relatedwork}

Mining the social web for crisis-relevant information has been an active and fruitful research topic for many years.
Our coverage of it focuses on overviewing methods for mining  (\S\ref{subsec:related-mining}), classification (\S\ref{subsec:related-classification}), and summarization (\S\ref{subsec:related-summarization}) of crisis-relevant social media messages.

\subsection{Mining Social Media for During Crises}
\label{subsec:related-mining}

Social media is a key communication channel during all kinds of crises, including natural and man-made disasters.
Computational methods from many disciplines can contribute to creating mining and retrieval systems that can help emergency managers \cite{castillo2016big}.
Crisis-related social spans many different categories of information,
including timely messages about urgent needs from affected populations and damaged infrastructure such as bridges or roads.
Together, this information is relevant for emergency response, recovery management, and assessments of the costs of damages \cite{doi:10.1177/0165551519828620}
Unfortunately, most methods for mining social media during disasters described in the extensive literature on the topic are monolingual, limiting the applicability in countries using languages other than English or even in English-speaking countries with increasing multilingual urban populations \cite{lorinisocial}.
The response during the disasters could be significantly improved with the ability to employ social data mining methods on user-generated data across multiple languages \cite{vitiugin2019comparison}.
Cross-lingual and multilingual classification and summarization methods provide an opportunity to gather complementary information across various languages spoken in affected areas.

\subsection{Classification of Crisis-Related Messages}
\label{subsec:related-classification}

In the recent literature on this topic, ``traditional'' supervised learning methods such as Naive Bayes and \ac{SVM} coexist with neural-network-based methods~\cite{saroj2020use}.
Indeed, \ac{SVM} for the classification crisis-relevant social media has consistently shown to exhibit high performance, especially when combined with semantic features computed with the help of external knowledge bases \cite{khare2018cross}.

Deep learning methods using various architectures have proven effective at detecting crisis-relevant messages; a popular architecture is Convolutional Neural Networks using word embeddings \cite{nguyen2017robust, lorini2019integrating}.
The addition of information specific to an event type, such as hydrological information in the case of floods, has been shown to improve classification performance \cite{de2020improving}.
A particularly influential model has been \ac{BERT} \cite{radford2018improving, devlin2018bert}, which is currently being used for various challenging \ac{NLP} tasks, including classification.
Recent papers on this topic describe end-to-end transformer-based models for crisis classification tasks, demonstrating promising results \cite{liu2021crisisbert, li2021combining}.

\ac{LASER} is an architecture to learn joint multilingual sentence representations for 93 languages.
The system uses a single \ac{BiLSTM} encoder with a shared byte-pair encoding vocabulary for all languages, coupled with an auxiliary decoder, and trained on publicly available parallel corpora.
The resulting embeddings are computed using English annotated data only and transferred to any of the 93 languages without any modification \cite{artetxe2019massively}.
LASER embeddings have also been shown to be effective in multilingual classification tasks \cite{pires2019multilingual, chi2021mt6}.

\subsection{Crisis-Related Information Summarization}
\label{subsec:related-summarization}

Social media messages are usually short and thus tend to provide fragmented information hence consolidating and summarizing information is key~ \cite{rudra2018identifying, rudra2018classifying}.
An informative summary can help stakeholders gain situational awareness and manage critical resources effectively~\cite{zade2018situational}.

The main approaches used for text summarization can be categorized as either extractive or abstractive~\cite{munot2014comparative}.
\emph{Extractive} approaches construct summaries by combining selected informative phrases or sometimes whole sentences from the source text \cite{dutta2019summarizing}.
\emph{Abstractive} summarization, on the other hand, generates summaries from a representation of the semantics of a given text;
an abstractive summary may contain words or sentences that do not appear in the source document(s).
Abstractive summarization techniques usually employ a generative approach \cite{lin2019abstractive,li2018guiding}.

Despite the benefits of abstractive approaches, extractive approaches are still considered state-of-the-art for summarization due to their simplicity and high performance~\cite{rossiello2017centroid, jadhav2018extractive}. 
However, extractive approaches often fail to include key elements useful in a report, such as answers to ``what,'' ``who,'' ``where,'' ``when,'' and ``how'' questions.
These are important elements in the domain of disaster and crisis management and need to be concisely incorporated into summaries \cite{kropczynski2018identifying}.
Query-based approaches have been described as a helpful manner of incorporating this information to improve the quality of reports~\cite{roush2020cx}.
In general, abstractive methods may facilitate the generation of more informative summaries, not restricted to sentences that directly take sequences of words from the source text~\cite{nafi2020abstractive}.
State-of-the-art abstractive summarization methods tend to adopt transformers and pre-trained models that have demonstrated great performance in other \ac{NLP} tasks: BART \cite{chen2021news}, T5 \cite{gupta2021automated}, PEGASUS \cite{sahu2021better}.

Our research builds upon previous work and contains two key innovations.
To the best of our knowledge, (1) we are the first to describe a summarization method that retrieves crisis-relevant information using a query-based approach, and (2) we are the first to propose a transformer-based summarization model for crisis-related messages.

% - - - - - - - - - - Method overview - - - - - - - - - -

\begin{figure*}
\includegraphics[width=1\textwidth]{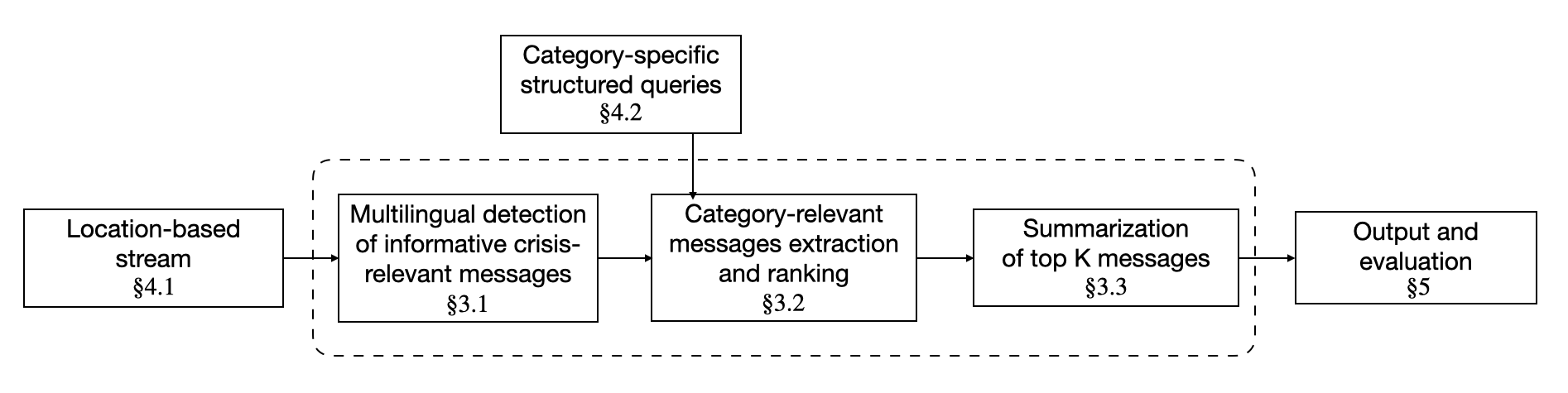}
\caption{Overview of our cross-lingual information summarization framework.}
\label{fig:framework}
\end{figure*}

\section{Method overview}
\label{sec:method}

In this section, we provide an overview of the proposed method, named \ourmethod (\ourmethodexpanded).
An overview of the method is shown in Figure~\ref{fig:framework}.
First, an automatic classification model is used for detecting crisis-relevant, informative messages (\S\ref{sec:model_class}).
Second, cross-lingual ranking is performed on these messages (\S\ref{sec:model_ranking}).
Third, the top $k$ ranked messages are given as input to a summarization model (\S\ref{sec:model_sum}).

\subsection{Classification Model}
\label{sec:model_class}

A large fraction of messages posted in social media in response to a crisis event doesn't include any informative claims beyond merely announcing that a crisis situation is developing.
Hence, a key step is detecting crisis-relevant \emph{informative} messages.
We model this as a binary classification task and create an automated classification model that we name \ourmethodclass (\ourmethodclassexpanded).

\spara{Pre-processing.}
Messages are preprocessed by replacing URLs and account mentions (``@user'') by specific tokens and turning hashtags into words.
We preserve punctuation and stopwords, as in the next steps, we use sentence embeddings and dependency parsing.

\spara{Feature extraction.}
We include morphological and syntactical features.
Using the Stanza \ac{POS} tagger \cite{qi2020stanza}, we count the number of numerals/numbers, nouns, verbs, adverbs, and adjectives in the messages.
Using the Stanza dependency parser, we extract and count syntactic features indicating the presence/absence of claim-containing sentences, such as subjective nouns, compounds, roots, and modality.
For both tasks, we use pre-trained models in each of the languages we work with.
We normalize the count of occurrences of each type of element in a message, such as ``contains  N numerals/numbers,'' ``contains N subjective nouns,'' and so on, using min-max scaling to be in $[0,1]$.
We also consider \ac{NER} features extracted using the SpaCy library \cite{honnibal2017spacy}.
These are binary features indicating whether a message contains persons' names, the name of a place, organization, or a date.
We use SpaCy's pre-trained models for each of the languages we work with. Off-the-shelf, SpaCy supports 15 languages, including all the ones we work with, except Croatian and Tagalog.
For Croatian, we use a contributed model for the Stanza package; for Tagalog, we use an open-source pre-trained model\footnote{https://github.com/matthewgo/FilipinoStanfordPOSTagger}.
Finally, we include \emph{message-specific features} indicating the (min-max scaled) number of URLs and user mentions in messages.

\spara{Embeddings.}
For representing the input data, we used sentence embeddings generated by the pre-trained transformer-based model, \ac{LASER} \cite{artetxe2019massively}.
\ac{LASER} sentence-level embeddings are obtained by applying max-pooling over the output of \iac{BiLSTM} sentence encoder. 
The \ac{BiLSTM} output is constructed by concatenating the outputs of two \acsp{LSTM} working in opposite directions (forward and backward). 
The bidirectional encoder captures more contextual information than a single-direction \ac{LSTM} encoder (e.g., a left-to-right one).
In our experiments, we used \ac{LASER} to embed all tweet sentences into fixed-size vectors of length 1,024.

\spara{Architecture.}
In our classification architecture, the embeddings are passed to an LSTM-layer and then combined with additional features.
This architecture is inspired by one proposed for the detection of fake news articles \cite{bhatt2017benefit}, which has also been used for emotion detection \cite{vitiugin2021emotion}.

The proposed scheme, depicted in Figure~\ref{fig:rank_architecture} (minus the query-based features, which are only used by the ranking step), computes the feature vectors separately and then combines them with the help of \iac{MLP} layer.
We use binary cross-entropy as the loss function to optimize and include a soft-max layer to classify social media text into one of two classes (``crisis relevant'' or ``not crisis relevant'').
The hyper-parameter settings of the feature extractor portions are shown in Table~\ref{table:hyperparams}.
The feature combination layer uses the softmax activation function with Adam optimizer, the learning rate of 0.001, batch size of 100, and binary cross-entropy loss.

\begin{table}
\small
\centering
\caption{Values of hyper-parameters.}
\label{table:hyperparams}
\begin{tabular}{p{0.2\textwidth}p{0.2\textwidth}p{0.2\textwidth}p{0.2\textwidth}}
%\begin{tabular}{l l l l}
 \toprule
 Hyperparameter & Text features & \acs{LASER} embeddings & Similarity features\\
 \midrule
 \acs{LSTM} layers & - & 1 & -\\
 \acs{MLP} layers & 2 & 3 & 2\\
 \acs{MLP} neurons & 128;24 & 1024;256;128 & 128;24\\
 Dropout & - & 0.5 & -\\
 Activation & relu & sigmoid & relu\\
 \bottomrule
\end{tabular}
%\vspace{-5mm}
\end{table}

\subsection{Cross-lingual Ranking Model}
\label{sec:model_ranking}

The next step in our method is to retrieve, from the informative crisis-related messages, a series of messages that are relevant for various informational categories.
Specifically, we retrieve and rank the top-$k$ most relevant messages from each category to pass them to the summarization model (\S\ref{sec:model_sum}).

To make information extraction more adaptable to different needs of emergency managers, our method is based on structured queries.
Each query is related to a specific information need and contains keywords, templates, and prototypes; a sample query is found in Table \ref{tab:query}.
\emph{Keywords} are words used frequently in messages of a category;
\emph{templates} are fragments containing key crisis-relevant facts; and
\emph{examples} are entire sentences or even entire messages corresponding to each category.

Each query is written in one language (English in our case) and used to extract information across all languages.
Queries are, for the most part, agnostic of the type of event, but in some cases, they may include elements that are specific to a type of disaster, for instance, in the case of earthquakes, their magnitude, or in the case of storms, rainfall.
We remark that context-based semantics allow our system to work even if these elements provided by the user are not 100\% complete (e.g., we can find messages containing related keywords or messages with similar semantics to the examples provided but using different wording).

\begin{figure*}
\centering
\includegraphics[width=0.65\textwidth]{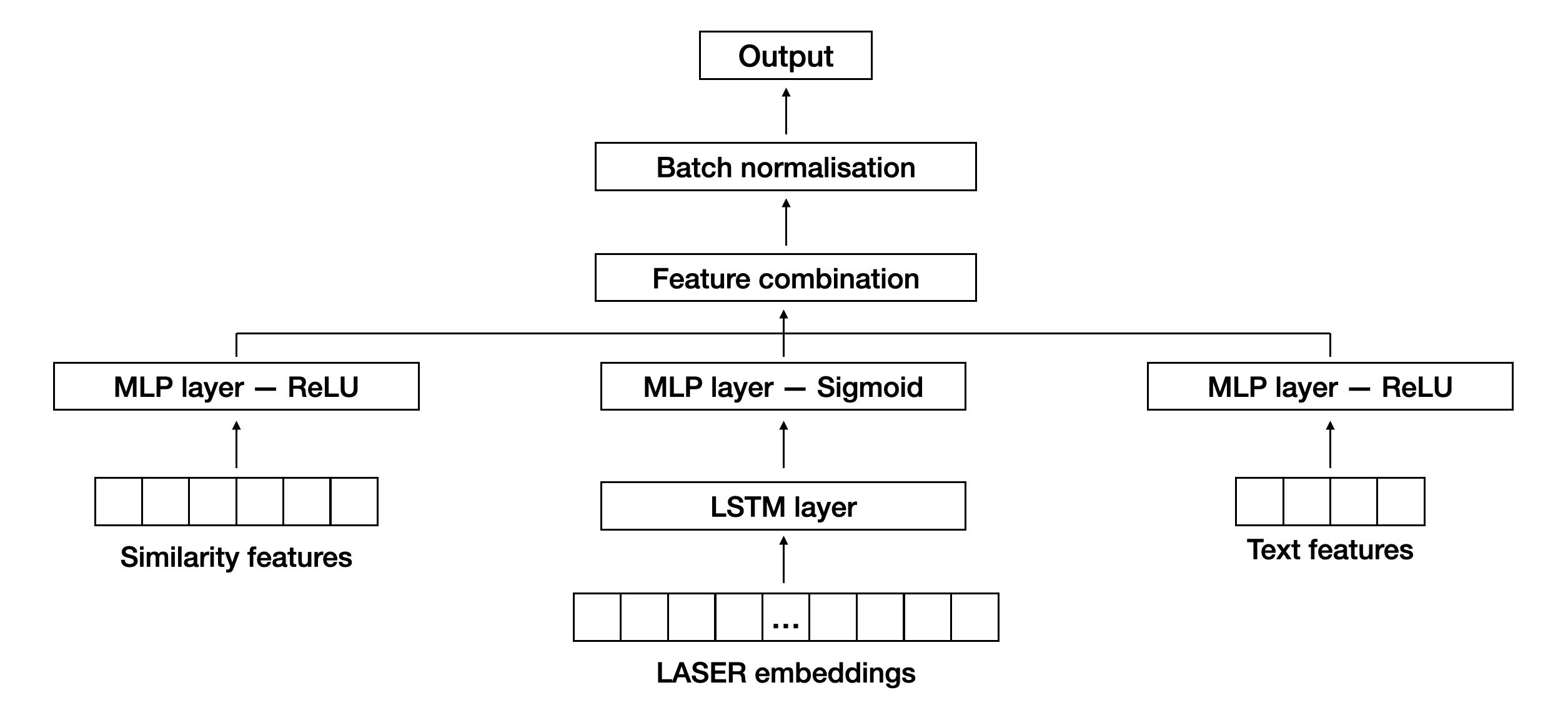}
\caption{Combining the transformer embeddings with morphological, syntactic, message-specific features and query similarity features using deep MLP}
\label{fig:rank_architecture}
\end{figure*}

To calculate query similarity features, we measure average and maximum cosine similarity between the query's keywords, templates, prototypes, and each message.
As a result, we have six similarity features.
For ranking messages, we use basically the same architecture as in the previous step (S\ref{sec:model_class}), with the addition of query similarity features.
This is depicted in Figure~\ref{fig:rank_architecture}.
After removing duplicates, we pass the top 100 candidates to the summarization step. We tested with the top 20, 50, and 100 candidates, and observed that the top 100 provided the highest recall.

\subsection{Summarization Model}
\label{sec:model_sum}

The final component of our method creates a category-specific summary from the retrieved messages for each category.
These summaries are created by T5, a pre-trained\footnote{HuggingFace - https://huggingface.co} transformer model widely used for summarization tasks~\cite{raffel2020exploring}.
In our preliminary experiments, this model performed better than a similarly pre-trained BART-based model.

We tested two different configurations for the summarization model: a regular condition and a diversified condition.
In the \emph{regular} condition, we gave T5 as input the texts from the top-100 most relevant candidates and limited the length of the output text.
In the \emph{diversified} condition, we clustered the top-100 most relevant candidates and gave to T5 as input the texts of all the messages on each cluster, one cluster at a time; the resulting summary is the concatenation of the per-cluster summaries.
For the diversification step, we first automatically find an appropriate number of clusters using the Silhouette Score.
Next, we cluster texts by the K-Means method.
Because we want to keep all summaries comparable in length for our experiments, we set a maximum number of clusters to four.
Also, during experiments, we found that the heuristic of summarizing clusters in decreasing order by size (i.e., starting with the largest cluster) helps to generate more relevant summaries.
This is akin to following the ``inverted pyramid'' style typically used in journalism.

% - - - - - - - - - - Experimental Setup and Evaluation - - - - - - - - - -

\section{Experimental Setup and Evaluation}
\label{sec:setup}

In this section, we describe our data collection (\S\ref{subsec:setup-collection}) and the queries used to retrieve relevant messages in various information categories (\S\ref{subsec:experiment-queries}). 
Then, we describe the baselines for classification (\S\ref{subsec:setup-categorization}) and summarization (\S\ref{subsec:setup-summarization}), and the evaluation metrics used to compare the proposed method against the baselines (\S\ref{subsec:setup-metrics}).

\subsection{Multilingual Data Collection}
\label{subsec:setup-collection}

\begin{table*}
\caption{Number of annotated messages for each event, including total number of messages, and number of messages labeled as informative by a human annotator. Local languages appear in bold.}
\small
\centering
\resizebox{\textwidth}{!}{\begin{tabular}{l l l l l l l l l l l l l }
\toprule
\multicolumn{2}{l}{} & \multicolumn{2}{l}{lang. 1 (en)} & \multicolumn{2}{l}{lang. 2 (es)} & \multicolumn{2}{l}{lang. 3 (fr)} & \multicolumn{2}{l}{language 4} & \multicolumn{2}{l}{language 5}\\
\textbf{} & dates & total & info & total & info & total & info & total & info & total & info\\\midrule
\textbf{Australia bushfires} & 06-31.01.2020 & \textbf{2000} & \textbf{233} & 2000 & 435 & 2000 & 460 & 2000 & 285 (ja) & 2000 & 167 (id)\\
\textbf{Fukushima earthquake} & 13.02.2021 & 2000 & 266 & 2000 & 529 & 2000 & 227 & \textbf{3000} & \textbf{101} (ja) & 3000 & 153 (id)\\
\textbf{Gloria storm} & 17-25.01.2020 & 703 & 393 & \textbf{571} & \textbf{210} & \textbf{517} & \textbf{168} & \textbf{542} & \textbf{233} (ca) & - & -\\
\textbf{Taal eruption} & 12-17.01.2020 & 551 & 123 & 691 & 202 & 610 & 114 & \textbf{1500} & \textbf{258} (tl) & 458 & 151 (pt)\\
\textbf{Zagreb earthquake} & 22-24.03.2020 & 537 & 162 & 509 & 243 & 520 & 187 & \textbf{1500} & \textbf{282} (hr) & 542 & 163 (de)\\\bottomrule
\end{tabular}}
\label{table:collection}
%\vspace{-5mm}
\end{table*}

Our data collection followed standard practices to collect crisis-related social media messages from Twitter.
We collected public tweets using Twitter’s public API, filtering by location-related keywords and date, without using any additional filtering (e.g., we did not restrict the query to specific languages). 
We considered five disaster events between January 2020 and February 2021 that received substantial news coverage internationally:
\begin{itemize}
    \item \emph{Australian bushfires} (2019-2020):  period of bushfires in many parts of Australia, which, due to its unusual intensity, size, duration, and uncontrollable dimension, was considered a ``megafire'';\footnote{https://en.wikipedia.org/wiki/2019–20\_Australian\_bushfire\_season}
    
    \item \emph{Fukushima earthquake} (February 2021): a 7.1 M\textsubscript{w} or 7.3 M\textsubscript{JMA} earthquake that struck offshore east of Tōhoku, Japan and caused significant structural damage across the Tōhoku and Kanto regions;\footnote{https://en.wikipedia.org/wiki/2021\_Fukushima\_earthquake}
    
    \item \emph{Gloria storm} (January 2020): a Mediterranean storm that affected eastern Spain and southern France with high winds and heavy rainfall;\footnote{https://en.wikipedia.org/wiki/Storm\_Gloria}
    
    \item \emph{Taal volcano eruption} (January 2020): a phreatomagmatic eruption from its main crater that spewed ashes across Calabarzon, Metro Manila, and some parts of Central Luzon and the Ilocos Region in the Philippines, resulting in temporary closures of schools and workplaces, and disruptions of flights in the area;\footnote{https://en.wikipedia.org/wiki/2020\_Taal\_Volcano\_eruption}
    
    \item \emph{Zagreb earthquake} (March 2020): an earthquake of magnitude 5.3 M\textsubscript{w}, 5.5 M\textsubscript{L}, which hit the capital of Croatia, causing severe damage to hundreds of buildings in its historical center.\footnote{https://en.wikipedia.org/wiki/2020\_Zagreb\_earthquake}
\end{itemize}
All messages include a ``language'' field computed by Twitter using a language detection model developed specifically for tweets.
We counted the number of messages per language in each event. 
Three of the top languages were common to all of the studied events: English (ISO 639-1 code: en), Spanish (es), and French (fr).
Additionally, we found several hundred messages for each event in other languages, including Catalan (ca), Tagalog (tl), Croatian (hr), German (de), Japanese (ja), Indonesian (id), and Portuguese (pt).
After collecting the data, we labelled tweets or their translation to English that contained potentially informative factual information. We name this group of tweets ``informative messages.''
One of the authors created the ground truth by reviewing each event and hand labeling these tweets.
Another author reviewed a portion of the classified tweets, and adjustments to the classification task were agreed upon when needed.
Additionally, ``informative messages'' were reviewed by crowdworkers during the categorization task and excluded if they did not contain information related to any category.
The number of annotated messages is shown in Table~\ref{table:collection}.

Next, we used crowdsourcing to further categorize the messages into various informational categories.
Specifically, we employed crowdworkers through a crowdsourcing platform,\footnote{SurgeHQ - https://www.surgehq.ai} paying the standard rate recommended by the platform.
We asked three different workers to label each of the approximately 5,700 informative messages across languages.
The target categories were based on an ontology from TREC-IS 2018 \cite{mccreadie2019trec}, where we grouped some low-level ontology categories into higher-level ones.
In total, we defined nine high-level classes of information, shown in Table \ref{table:ontology}.

\begin{table*}
\caption{Categories for multilingual information extraction, based on the ontology from TREC-IS 2018~\cite{mccreadie2019trec}. Example messages have been paraphrased for anonymity.}
\small
\centering
\begin{tabular}{l l l}\toprule
 \textbf{Category} & \textbf{Description} & \textbf{Example message} \\
 \midrule
 %Blackout & Power and communication outages & \textit{Storm leaving 200 without electricity} \\
 Casualties & Affected or injured people & \textit{Around 150 injured people} \\
 Damage & Built or natural environment damage & \textit{Destroying orange trees and rice paddies} \\
 Danger & Messages of caution or alerts & \textit{RED WARNING Barcelona - Danger to life} \\
 Government & Official report by public agencies & \textit{Local authorities continuing the search for ...} \\
 Sensor & Seismic activity & \textit{Zargeb hit by 5.3 magnitude earthquake} \\
 Service & Providing a service or help & \textit{Local org. provides shelter for more than 1,000 people} \\
 Water & Water-related messages & \textit{floods in Catalonia} \\
 Weather & Weather updates & \textit{heavy rainfall and flooding across region} \\\bottomrule
\end{tabular}
\label{table:ontology}
%\vspace{-5mm}
\end{table*}

\subsection{Queries}
\label{subsec:experiment-queries}

We use a set of queries covering the nine information categories listed in Table~\ref{table:ontology}.
As described in \S\ref{sec:model_ranking}, a query for an information category includes keywords, templates, and prototypes.
Creating a query requires some degree of familiarity with social media messages posted during emergency situations.

\emph{Keywords} are nouns and verbs usually present in messages containing a specific category of information.
Practitioners could complete this task with scripts or programs to find frequent words or phrases present in previous collections of messages from past events.
\emph{Templates} are small fragments of text describing crisis-relevant facts.
The kind of information that we seek is in situation reports %such as \emph{ERCC text reports},
or in Wikipedia disaster-related \emph{infoboxes}, which are templates that Wikipedia editors use to summarize crisis information.
Users can provide such templates by copy-pasting passages from these sources, replacing the numbers or locations found there with the tokens \texttt{NUMBER} or \texttt{LOCATION}.
Finally, users can provide \emph{prototypes} -- example messages or central passages typical in category-related texts, which can be obtained by sampling diverse, informative messages from past events.
We envision a specialized user interface may assist users in formulating such queries, and we plan to explore that in future work.
The scope of this paper is to demonstrate the approach and provide an initial set of easily extended and refined queries.
One such query is shown in Table~\ref{tab:query}.

\begin{table}[ht]
\caption{Example query. Each query includes keywords, templates, and prototypes.}
\small
\begin{tabular}{l l}
\toprule
\multicolumn{2}{c}{Query for category: Weather} \\
\midrule
keywords: & \textit{snow, weather, rain, wind, coast, mph, kmh, forecast} \\ 
templates: & \textit{batter parts of LOC, damages from winds, pummelling the region, NUMBER km/h winds,} \\
& \textit{weather forecast, bad weather, heavy snow, strong wind, storm is hitting, wind gust} \\ 
prototypes: & \textit{Wind, rain and snow batter parts of country} \\ 
& \textit{Storm brought around NUMBER m of snow and affect rivers} \\ 
& \textit{Heavy rainfall, strong wind and more than NUMBER of snow across LOC} \\
& \textit{Storm is hitting eastern LOC, with high winds and heavy rain} \\
& \textit{Storm has battered parts of LOC and reportedly brought worth of rain} \\
& \textit{Maximum gusts of wind in LOC NUMBER km / h} \\
& \textit{Tonight, terrible rains in LOC} \\
& \textit{Organisation has so far done NUM health care due to strong winds} \\
& \textit{Gusts of wind left fallen trees} \\
\bottomrule
\label{tab:query}
\end{tabular}
%\vspace{-5mm}
\end{table}

\subsection{Message Classification Schemes}
\label{subsec:setup-categorization}

To compare our proposed method for informative messages detection, we construct baseline models using one classical machine learning scheme (\acs{SVM}) and one deep learning scheme (\acs{LSTM}) that uses \ac{LASER} embeddings as input features.
Also, we compare our model with a cross-lingual LinearSVC-based model that uses semantic features extracted with the BabelNet knowledge base\footnote{https://babelnet.org}.
The complete list of proposed modeling schemes for evaluation is the following:

\begin{itemize}
    \item \textbf{LASER+SVM}: this method uses pre-trained LASER embeddings; the embeddings are then classified by a Linear \ac{SVM} model;
    \item \textbf{LASER+LSTM}: this method uses pre-trained \ac{LASER} embeddings; the embeddings are then classified by a \ac{LSTM} model;
    \item \textbf{Khare}~\cite{khare2018cross}: this is a cross-lingual classification approach that uses additional semantic features extracted from external knowledge bases; % (the method is reported as ``SF+SemBNDB'' in their paper)
    \item \textbf{CrisisBERT}~\cite{liu2021crisisbert}: this is an end-to-end transformer-based model for crisis classification tasks (our implementation uses the DistilBERT~\cite{sanh2019distilbert} architecture);
    \item \textbf{\ourmethodclass} (ours): this is our method for classification, using a combination of \acs{LASER} embeddings and tweet-related features. classified by a \ac{LSTM} model.
\end{itemize}

\subsection{Summarization Methods}
\label{subsec:setup-summarization}

We compare our \textit{\ourmethod} model and its diversified variation \textit{\ourmethoddiv} against several state-of-the-art summarization models.
With the exception of the \textit{LASER+LSTM+T5} method, all of the baselines use only category-related tweets as input, i.e., we simulate the best scenario in which the input is received from a perfect classifier.
In our proposed models, we use the query-based model we described.
The complete list of baselines for summarization that we used is the following:

\begin{itemize}
    \item \textbf{LASER+LSTM+T5}: this method uses pre-trained \ac{LASER} embeddings,  which are passed as input to a \ac{LSTM} model for category classification and then to a T5 model for summarization;
    \item \textbf{C-SKIP}~\cite{rossiello2017centroid}: this is a centroid-based method using a FastText skipgram model trained on the CrisisLexT26 dataset~\cite{olteanu2014crisislex}, improved by the use of T5 pre-trained model (originally, the method used a corpus extracted from  Google News);
    \item \textbf{CX\_DB8}~\cite{roush2020cx}: this is a queryable word-level unsupervised extractive summarizer, which is based on the text embedding framework Flair~\cite{akbik2018contextual}.
    We tested this with different pre-trained embeddings, including transformer-based such as \ac{BERT} and XLNet; for this task and datasets, the best results were obtained with \ac{GLOVE} embeddings;
    \item \textbf{NAFI}~\cite{nafi2020abstractive}: this is an abstractive text summarization method developed specifically for crisis events;
    \item \textbf{\ourmethod} (ours): we use a combination of \acs{LASER} embeddings with tweet-related features and query similarities features that are passed to a \acs{LSTM} model for the ranking step and then uses a T5 model for the summarization step;
    \item \textbf{\ourmethoddiv} (ours): this is the same as \ourmethod but retrieves diversified (see \S\ref{sec:model_sum}) top-k candidates in the ranking step.
\end{itemize}

\subsection{Evaluation Metrics}
\label{subsec:setup-metrics}

To evaluate the performance of the \emph{classification} models, we use three standard metrics: Accuracy (ACC), Area Under the Receiver Operating Characteristic Curve (AUC), and weighted F-measure (F1).
These metrics are typically used in research on social media for emergency management (e.g., \cite{khare2018cross, lorini2019integrating, vitiugin2019comparison}).

To evaluate the \emph{summarization} models, we considered four methods.
First, we annotated all summaries for factual claims and then computed, for each summary, the fraction of factual claims it contained out of the total factual claims mentioned across all summaries.
Second, we computed the BERTScore~\cite{zhang2019bertscore} of each summary, which is a metric for evaluation of a text that compares them against a ground truth; in our case, an official report about the event.
Third, we performed a crowdsourced evaluation of the readability of each summary across five dimensions: grammaticality, non-redundancy, referential clarity, focus, and structure and coherence \cite{iskender2019crowdsourcing}; five crowdsourcing workers were asked to compare summaries across each dimension.
Fourth, we asked three experts in emergency management to perform a side-by-side comparison of the summaries and computed the number of times each summary was preferred.

% - - - - - - - - - - Results - - - - - - - - - -

\section{Results}
\label{sec:results}

In this section, we present the results of our evaluation and comparison with state-of-the-art methods.
First, we present an evaluation of our classification method (\S\ref{subsec:results-classification}).
Next, we consider the extent to which summaries are comprehensive in terms of factual claims (\S\ref{subsec:results-factual}).
Then, we ask crowdsourcing workers to evaluate the readability of summaries (\S\ref{subsec:results-bertscore}).
Finally, we ask experts to perform a side-by-side comparison of the summaries (\S\ref{subsec:results-expert}).

\subsection{Cross-lingual Classification}
\label{subsec:results-classification}

The first experiment is a ``leave-one-language-out'' evaluation: for each event, the classifier is trained on data from 3 or 4 languages and tested on the last language.
What we simulate here is a scenario in which we have labeled data in several languages and extract information in a new language.
Table \ref{table:classification} shows the performance comparison of our method \textit{\ourmethodclass} (\ourmethodclassexpanded) with other baselines and state-of-the-art.
We also perform a ``leave-one-event-out'' evaluation, in which we train on multilingual data for all events except one and test on the event that was left out.
Results are shown on Table~\ref{table:crisis-out}.

We can observe that in general methods based on multilingual transformers perform better than the semantic-based model by \textit{Khare}. 
However, there are a few differences between schemes based on \ac{LASER} embeddings; \ac{SVM} and \ac{LSTM} achieve in general the best performance, with some variations across datasets.
The performance of \textit{CrisisBERT} is comparable to that of the proposed method \textit{\ourmethodclass} in some cases and across some metrics, but the average performance of \textit{\ourmethodclass} is better.
We also observe that across all methods, ``leave-one-event-out'' seems to pose a more difficult problem than ``leave-one-language-out.''
This suggests that most of the multilingual methods we tested capture fairly well event-specific concepts (such as specific places, impacts, or needs) of each crisis but do not generalize so well across events.

\begin{table*}[ht]
\caption{Results of cross-validation evaluation of message classification across languages (``leave-one-language-out''), with 4-5 languages per event: the test data contains all of the messages for an event in one language, while the training data contains messages in other language for the same event.}
\small
\centering
\resizebox{\textwidth}{!}
{\begin{tabular}{l|ccc|ccc|ccc|ccc|ccc|ccc}
\toprule
 & \multicolumn{3}{|c}{Australia bushfires} & \multicolumn{3}{|c}{Fukushima earthquake} & \multicolumn{3}{|c}{Gloria storm} & \multicolumn{3}{|c}{Taal eruption} & \multicolumn{3}{|c}{Zagreb earthquake} & \multicolumn{3}{|c}{Average}\\
Schemes & ACC & F1 & AUC & ACC & F1 & AUC & ACC & F1 & AUC & ACC & F1 & AUC & ACC & F1 & AUC & ACC & F1 & AUC\\\midrule
\textbf{LASER+SVM} & 92.3 & 91.1 & 70.7 & 97.4 & \textbf{97.3} & 85.4 & 83.6 & 81.6 & 66.6 & 95.4 & \textbf{94.4} & 77.9 & 92.9 & \textbf{92.9} & 90.6 & 92.3 & 91.5 & 78.2\\
\textbf{LASER+LSTM} & 91.9 & 74.6 & 92.7 & 97.5 & 92.7 & 99.5 & 83.2 & 66.9 & 86.2 & 94.5 & 84.3 & 96.7 & 93.7 & 83.0 & 97.2 & 92.2 & 80.3 & 94.4\\
\textbf{Khare} & 88.8 & 85.8 & 64.1 & 88.1 & 85.3 & 65.9 & 56.3 & 50.1 & 70.5 & 90.6 & 86.1 & 50.0 & 41.6 & 30.2 & 33.3 & 73.1 & 67.5 & 58.0\\
\textbf{CrisisBERT} & 91.4 & 87.1 & 96.2 & \textbf{97.6} & 96.6 & \textbf{99.4} & 87.7 & 83.4 & 94.2 & 95.3 & 92.6 & 98.0 & \textbf{94.1} & 90.6 & 98.3 & 93.2 & 90.1 & 97.2\\
\textbf{*\ourmethodclass} & \textbf{95.9} & \textbf{93.6} & \textbf{99.2} & \textbf{97.6} & 96.3 & \textbf{99.4} & \textbf{93.0} & \textbf{90.9} & \textbf{97.8} & \textbf{95.6} & 93.3 & \textbf{98.1} & 93.2 & 88.5 & \textbf{98.6} & \textbf{95.1} & \textbf{92.5} & \textbf{98.6}\\\bottomrule
\end{tabular}}
\label{table:classification}
%\vspace{-3mm}
\end{table*}

\begin{table*}[ht]
\caption{Results of cross-validation evaluation of message classification across events (``leave-one-event-out''): the test data contains all of the messages for one event, while the training data contains messages from all of the other events.}
\small
\centering
\resizebox{\textwidth}{!}{
\begin{tabular}{l|ccc|ccc|ccc|ccc|ccc|ccc}
\toprule
 & \multicolumn{3}{|c}{Australia bushfires} & \multicolumn{3}{|c}{Fukushima earthquake} & \multicolumn{3}{|c}{Gloria storm} & \multicolumn{3}{|c}{Taal eruption} & \multicolumn{3}{|c}{Zagreb earthquake} & \multicolumn{3}{|c}{Average}\\
Schemes & ACC & F1 & AUC & ACC & F1 & AUC & ACC & F1 & AUC & ACC & F1 & AUC & ACC & F1 & AUC & ACC & F1 & AUC\\\midrule
\textbf{LASER+SVM} & \textbf{90.1} & 87.8 & 63.0 & 96.4 & \textbf{96.2} & 84.8 & \textbf{87.3} & \textbf{85.8} & 71.4 & 93.1 & 91.6 & 65.6 & 85.7 & 84.0 & 74.0 & 90.5 & 89.1 & 71.8\\
\textbf{LASER+LSTM} & 87.5 & 66.4 & 87.5 & 96.6 & 83.4 & 99.0 & 80.3 & 63.5 & 80.3 & 93.6 & 81.4 & 96.9 & 84.2 & 75.4 & 92.5 & 88.4 & 74.0 & 91.2\\
\textbf{Khare} & 86.2 & 83.3 & 54.8 & 93.4 & 93.4 & 81.9 & 80.8 & 77.1 & 57.9 & 90.2 & 88.4 & 58.8 & 80.6 & 77.4 & 65.9 & 86.2 & 83.9 & 63.9\\
\textbf{CrisisBERT} & 88.0 & 88.0 & 93.3 & \textbf{96.7} & 95.9 & \textbf{99.3} & 83.3 & 82.2 & 91.7 & \textbf{94.6} & 93.4 & \textbf{98.0} & 87.0 & 82.1 & 94.4 & 89.9  & 	88.3 & 95.3\\
\textbf{*\ourmethodclass} & 89.1 & \textbf{88.2} & \textbf{94.0} & 96.1 & 95.9 & 98.4 & 86.4 & 84.8 & \textbf{93.3} & 93.7 & \textbf{93.5} & 96.3 & \textbf{88.4} & \textbf{84.4} & \textbf{95.0} & \textbf{90.7} & \textbf{89.4} & \textbf{95.4}\\\bottomrule
\end{tabular}}
\label{table:crisis-out}
%\vspace{-3mm}
\end{table*}

\subsection{Recall of Factual Claims}
\label{subsec:results-factual}

One way of measuring how informative are different summaries are, is to consider the extent to which they contain factual claims related to an information category for an event.
To perform this evaluation, we manually coded each category-related factual claim in each of the generated summaries across all methods and then counted the number of claims in every summary compared to the overall claims.\footnote{These annotated summaries are part of our data release.}
The fraction of claims contained in summary is divided by the total number of claims across all summaries if what we call the \emph{recall of factual claims.}
Table \ref{table:sum_recall} shows the performance comparison of \textit{\ourmethod} with other baselines and state-of-the-art.
\textit{\ourmethod} and \textit{C-SKIP} outperform the other methods in both the cross-lingual and English-only evaluation, with a small advantage for \textit{\ourmethod}.

The diversified method \textit{\ourmethoddiv} produces summaries with less factual claims than \textit{\ourmethod}; in our observations, this is partially explained by diversification leading to more low-ranking claims to be included, and those claims are more likely to be incorrectly associated to the category under analysis.
In other words, the lower we go on the list of retrieved messages for a category, the more likely we are to find messages that actually belong to other categories.
As we explain in subsection \ref{sec:method}, we create clusters from top-100 candidates, and often the number of category-related candidates is much less than 100; in this case, clusters other than the first one are likely to be non-related to a category.

\begin{table}[hb]
\caption{Factual claims present in each summary, on average, as a percentage of the total number of factual claims across all summaries for a category and event. We consider cross-lingual and English-only evaluations. The top two methods are extractive, while the remaining four are abstractive; our methods are marked with an asterisk.}
\small
\centering
\begin{tabular}{l c c } \toprule
Scheme & Cross-lingual & English \\\midrule
\textbf{C-SKIP} & 28.8\% & 29.0\% \\
\textbf{CD\_DB8} & 24.1\% & 17.8\% \\ \midrule
\textbf{LASER+LSTM+T5} & 10.0\% & 16.0\% \\
\textbf{Nafi} & 25.1\% & 28.8\% \\
\textbf{*\ourmethod} & \textbf{29.4\%} & \textbf{31.2\%}\\
\textbf{*\ourmethoddiv} & 23.8\% & 23.5\%\\
\bottomrule
\end{tabular}
\label{table:sum_recall}
%\vspace{-5mm}
\end{table}

\subsection{BERTScore - Similarity with Official Reports}
\label{subsec:results-bertscore}

To perform this evaluation, we retrieve summaries for each event prepared by the \ac{ERCC}.\footnote{ERCC Portal - https://erccportal.jrc.ec.europa.eu/}
These summaries are created at the level of entire events and not divided by category. Hence, we create an event-level summary using each method by combining the summaries from all categories.
Each event-level summary is compared against the \ac{ERCC} one using BERTScore.

Table \ref{table:bertscore} shows the results of the performance comparison of \textit{\ourmethod} against other models.
Both proposed methods \textit{\ourmethod} and \textit{\ourmethoddiv} show better performance than the baselines in all datasets except one.
The exception is \textit{C-SKIP}, a centroid-based extractive method, which demonstrates higher performance for one of the analyzed events (Zagreb earthquake).
The BERTScore evaluation also helps us interpret the results regarding the recall of factual claims, as often the precision of \textit{\ourmethod} is higher than the one of \textit{\ourmethoddiv}.

\begin{table*}
\caption{Comparison of cross-lingual summaries against reports by 
\acs{ERCC} using BERTScore: precision (P), Recall (R), and F1 measure (F1). The top two methods are extractive, while the remaining four are abstractive; ours are marked with an asterisk.}
\small
\centering
\resizebox{\textwidth}{!}
{\begin{tabular}{l|ccc|ccc|ccc|ccc|ccc|ccc}
\toprule
 & \multicolumn{3}{|c}{Australia bushfires} & \multicolumn{3}{|c}{Fukushima earthquake} & \multicolumn{3}{|c}{Gloria storm} & \multicolumn{3}{|c}{Taal eruption} & \multicolumn{3}{|c}{Zagreb earthquake} & \multicolumn{3}{|c}{Average}\\
Schemes 
 & P & R & F1 
 & P & R & F1  
 & P & R & F1 
 & P & R & F1 
 & P & R & F1 
 & P & R & F1 \\\midrule
\textbf{C-SKIP} & 78.9 & 79.5 & 79.2 & 79.9 & 84.0 & 81.9 & 79.5 & 81.5 & 80.5 & 80.6 & 80.7 & 80.7 & \textbf{82.5} & \textbf{82.8} & \textbf{82.6} & 80.3 & 81.7 & 81.0\\
\textbf{CX\_DB8} & 75.2 & 77.4 & 76.3 & 78.5 & 80.8 & 79.6 & 76.3 & 78.1 & 77.2 & 77.3 & 78.8 & 78.0 & 77.6 & 78.9 & 78.2 & 77.0 & 78.8 & 77.9\\\midrule
\textbf{LSTM+LASER+T5} & 79.5 & 78.6 & 79.0 & 81.8 & 83.8 & 82.8 & 79.9 & 81.5 & 80.7 & 75.5 & 78.3 & 76.9 & 82.3 & 81.9 & 82.1 & 79.8 & 80.8 & 80.3\\
\textbf{Nafi} & 76.0 & 78.6 & 77.3 & 78.9 & 82.1 & 80.5 & 77.6 & 79.8 & 78.7 & 77.0 & 79.8 & 78.4 & 80.3 & 82.4 & 81.3 & 78.0 & 80.5 & 79.2\\
\textbf{*\ourmethod} & \textbf{81.0} & \textbf{80.6} & \textbf{80.8} & 80.6 & \textbf{85.4} & 82.9 & \textbf{81.2} & 82.0 & \textbf{81.6} & \textbf{82.7} & \textbf{81.2} & \textbf{81.9} & 81.8 & 82.2 & 82.0 & \textbf{81.5} & \textbf{82.3} & \textbf{81.9}\\
\textbf{*\ourmethoddiv} & 80.5 & 80.1 & 80.3 & \textbf{83.1} & 84.8 & \textbf{83.9} & 80.9 & \textbf{82.2} & 81.5 & 82.4 & 81.1 & 81.7 & 80.6 & 82.6 & 81.6 & \textbf{81.5} & 82.2 & 81.8\\\bottomrule
\end{tabular}}
\label{table:bertscore}
%\vspace{-2mm}
\end{table*}

Additionally, the comparison of \textit{\ourmethod} against other models, considering only English messages as input for all models, is presented in Table \ref{table:en-bertscore}.
In this monolingual evaluation, methods are closer to each other in terms of BERTscore similarity with the reference.
\textit{\ourmethoddiv} models show slightly better average performance.
In the Australia bushfires dataset, \textit{C-SKIP} performs better, and in the Taal volcano eruption dataset \textit{\ourmethod} performs slightly better.

\begin{table*}[hb]
\caption{Comparison of English-only summaries against reports by 
\acs{ERCC} using BERTScore: precision (P), Recall (R), and F1 measure (F1). The top two methods are extractive, while the remaining four are abstractive; ours are marked with an asterisk.}
\small
\centering
\resizebox{\textwidth}{!}
{\begin{tabular}{l|ccc|ccc|ccc|ccc|ccc|ccc}
\toprule
 & \multicolumn{3}{|c}{Australia bushfires} & \multicolumn{3}{|c}{Fukushima earthquake} & \multicolumn{3}{|c}{Gloria storm} & \multicolumn{3}{|c}{Taal eruption} & \multicolumn{3}{|c}{Zagreb earthquake} & \multicolumn{3}{|c}{Average}\\
Schemes 
 & P & R & F1
 & P & R & F1
 & P & R & F1
 & P & R & F1
 & P & R & F1
 & P & R & F1\\\midrule
\textbf{C-SKIP} & \textbf{78.2} & \textbf{79.2} & \textbf{78.7} & 79.8 & 80.7 & 80.3 & 77.4 & 78.9 & 78.2 & 77.9 & 80.1 & 79.0 & 80.2 & 81.0 & 80.6 & 78.7 & 80.0 & 79.3\\
\textbf{CX\_DB8} & 75.8 & 77.2 & 76.5 & 79.9 & 79.2 & 79.5 & 76.1 & 77.1 & 76.6 & 75.7 & 77.6 & 76.7 & 79.2 & 78.9 & 79.0 & 77.3 & 78.0 & 77.7\\
\midrule
\textbf{LSTM+LASER+T5} & 75.9 & 78.1 & 77.0 & \textbf{81.9} & 80.8 & 81.3 & 78.0 & 79.1 & 78.5 & \textbf{78.0} & 79.7 & 78.9 & 81.4 & 81.3 & 81.3 & 79.0 & 79.8 & 79.4\\
\textbf{Nafi} & 75.0 & 78.2 & 76.6 & 77.6 & 80.9 & 79.2 & 75.3 & 77.8 & 76.6 & 72.9 & 78.0 & 75.4 & 77.5 & 79.7 & 78.6 & 75.7 & 78.9 & 77.3\\
\textbf{*\ourmethod} & 77.9 & 78.8 & 78.3 & 80.4 & 80.7 & 80.5 & 77.6 & 79.4 & 78.5 & 77.9 & \textbf{80.2} & \textbf{79.0} & 80.8 & 82.2 & 81.5 & 78.9 & 80.3 & 79.6\\
\textbf{*\ourmethoddiv} & 78.2 & 78.8 & 78.5 & 81.7 & \textbf{81.4} & \textbf{81.6} & \textbf{79.0} & \textbf{79.8} & \textbf{79.4} & 77.7 & 80.0 & 78.8 & \textbf{81.6} & \textbf{82.5} & \textbf{82.1} & \textbf{79.6} & \textbf{80.5} & \textbf{80.1}\\\bottomrule
\end{tabular}}
\label{table:en-bertscore}
%\vspace{-5mm}
\end{table*}

\subsection{Readability evaluation}
\label{subsec:results-readability}

The \emph{readability} of crisis reports is crucial to provide information to practitioners in an understandable way \cite{temnikova2015case}.
We considered five dimensions of readability: grammaticality, non-redundancy, referential clarity, focus, structure, and coherence \cite{iskender2019crowdsourcing}.
We performed this evaluation through a crowdsourcing platform,\footnote{SurgeHQ - https://www.surgehq.ai/} and computed our results by aggregating the assessments of five different annotators.
Annotators were shown an explanation of each annotation dimension before starting the evaluation.

A total of 43 evaluation rounds were performed, and in each round, the five annotators were shown independently the six summaries in random ordering.
They were asked to pick one of them as the best in terms of each evaluation dimension.
Then, we computed the best method for each round by majority voting among the five annotators.
The results of the evaluation, shown in Table~\ref{table:crowdworkers}, indicate that annotators considered summaries generated by \textit{\ourmethoddiv} as less redundant, more referentially clear, more focused, and more structured and coherent than the summaries generated by other methods.
On the other hand, \textit{C-SKIP} summaries, which are extractive summaries generated by a centroid-based method, were considered as more grammatically correct.

\begin{table*}
\caption{Readability evaluation of summaries, expressed as the percentage of times a method was chosen as the best for a given  dimension (column). The top two methods are extractive, the remaining four are abstractive; ours are marked with an asterisk.}
\small
\centering
\begin{tabular}{l r r r r r} \toprule
 & 
 & \multicolumn{1}{c}{Non-}
 & \multicolumn{1}{c}{Referential}
 & 
 & \multicolumn{1}{c}{Structure and}
\\
Scheme 
 & \multicolumn{1}{c}{Grammaticality}
 & \multicolumn{1}{c}{Redundancy} 
 & \multicolumn{1}{c}{Clarity} 
 & \multicolumn{1}{c}{Focus}
 & \multicolumn{1}{c}{Coherence} \\
 \midrule
 \textbf{C-SKIP} & \textbf{33.5\%} & 14.9\% & 14.9\% & 10.2\% & 13.0\% \\
 \textbf{CD\_DB8} & 0.5\% & 4.2\% & 2.8\% & 2.8\% & 1.4\% \\
 \midrule
 \textbf{LSTM+LASER+T5} & 19.5\% & 25.6\% & 20.0\% & 23.7\% & 25.6\% \\
 \textbf{Nafi} & 1.4\% & 1.4\% & 5.6\% & 4.2\% & 1.4\% \\
 \textbf{*\ourmethod} & 20.0\% & 20.9\% & 27.0\% & 24.7\% & 24.2\%\\
 \textbf{*\ourmethoddiv} & 25.1\% & \textbf{33.0\%} & \textbf{29.8\%} & \textbf{34.4\%} & \textbf{34.4\%}\\
  \bottomrule
\end{tabular}
\label{table:crowdworkers}
%\vspace{-2mm}
\end{table*}

\subsection{Expert Evaluation}
\label{subsec:results-expert}

The last evaluation involved three experts in emergency management, none of them a co-author of this paper, working in three different EU countries: (1) an operations coordinator with a VOST organization, (2) a program manager at a Civil Protection Department, and (3) a project manager at an Emergency Management System.
Experts were shown 43 pairs of summaries randomly selected from the five events; in each pair, which one of the summaries was generated by our method and the other by one of the baselines.
Each summary was accompanied by references (links) to source tweets related to each sentence or passage in the summary, as in the following example, in which underlined letters represent links:

\begin{quote}
\textit{at least 20,000 people have taken refuge in evacuation centers. evacuees need masks. there are also many evacuees in need. evacuees need food, water, shelter and medical help.[\underline{a}] more than 30 thousand people evacuated due to the eruption of the Taal volcano in the Philippines. Taal volcano eruption threatens the lives of more than 900,000 inhabitants. eruption of the Taal volcano in the Philippines has caused more than 24,000 people to be evacuated. [\underline{b},\underline{c},\underline{d}]}
\end{quote}

The evaluation was performed through an online form and was ``blind'' in the sense that the experts did not know, in each pair, which summary was generated by which method; the ordering of each pair was randomly chosen.
We performed two evaluation rounds, the first one comparing \textit{\ourmethoddiv} against \textit{Nafi}, and the second one comparing \textit{\ourmethoddiv} against \textit{LSTM+LASER+T5}.
Experts were asked to chose in a 5-points scale whether they (1) preferred summary 1, (2) had a slight preference for summary 1, (3) considered both summaries equally preferable, (4) had a slight preference for summary 2, or (5) preferred summary 2.
We additionally asked respondents to optionally comment on the quality of both summaries, if considered appropriate.

\begin{table}
\caption{Expert evaluation results: percentage of answers received, aggregated across three experts.}
\small
\centering
\begin{tabular}{rcccccl}\toprule
           & $\leftarrow$ Prefers & Prefers slightly & Both equal & Prefers slightly & Prefers $\rightarrow$ & \\ \midrule
\textbf{*\ourmethod} & 63.6\% & 14.7\% & 5.4\% & 8.5\% & 7.8\% & \textbf{Nafi} \\
\textbf{*\ourmethod} & 55.8\% & 14.0\% & 12.4\% & 9.3\% & 8.5\% & \textbf{LSTM+LASER+T5}\\
\bottomrule
\end{tabular}
\label{table:practitioners}
%\vspace{-5mm}
\end{table}

The results of the evaluation, shown in Table \ref{table:practitioners}, indicate that the consulted practitioners clearly preferred summaries generated by the proposed method in 64\% of the cases when compared with \textit{Nafi}, and in 56\% of the cases when compared with \textit{LSTM+LASER+T5}.
Practitioners' opinions about summaries mentioned that \textit{\ourmethoddiv} summaries were ``more accurate and with fewer repetitions,'' ``more information and better organized,'' ``better explained,''``more understandable,'' and they contain ``fewer mistakes and more data.''
In comparison, according to their comments, they tended to reject summaries that contain ``contradictory information,'' ``a lot of repetitions,'' ``more mistakes and subjects mixed.''
Inspecting the evaluation, we noticed that in general \textit{\ourmethoddiv's} had lower performance in summaries related to the \textit{Zagreb earthquake} event.
In contrast, in other events, our method was often preferred.
%
% WE DID NOT EXPLAIN ANYTHING ABOUT SOME SUMMARIES BEING ENGLISH-ONLY AND OTHERS BEING CROSS-LINGUAL -- CHATO
%Compared to \textit{Nafi}, our approach is more preferable in English-only summaries than cross-lingual. However, in the case of corresponding to \textit{LSTM+LASER+T5}, cross-lingual summaries are more preferable to English-only.

% - - - - - - - - - - Experimental Setup and Evaluation - - - - - - - - - -

\section{Conclusions, Limitations, and Future Work}
\label{sec:conclusions}

We have described a method for generating informative reports about crises from multilingual social media.
This method is based on structured queries, which are matched against messages that potentially contain the information we are interested on.
Queries are straightforward to construct, which means this method can be extended to a large variety of information needs.
Experiments with five different disaster events indicate that we can generate high-quality, readable reports from the messages and that practitioners might prefer the summaries generated by \textit{\ourmethoddiv} to those generated using state-of-the-art methods.

\smallskip\noindent\textbf{Limitations.}
The applicability of our method depends on some resources that are multilingual but that do not cover all existing languages.
Currently, we use Stanza for POS tagging, and dependency parsing \cite{qi2020stanza} (50 languages available through community contributions) and LASER embeddings to represent sentences (93 languages available).
Our method only works with 50 languages in the intersection of those supported by the POS tagger, dependency parser, and embeddings.
Using the proposed method for unsupported languages may require training new models for the additional languages.
In the work, we generated only English summaries which were useful for practitioners' evaluation. The generation of summaries in other languages could show different results.
The proposed approach is flexible and allows including additional categories of information with help of queries, but we have not tested that at this point.
Finally, the use of sentence embeddings (LASER) allows using the same approach for other social media (Facebook, Reddit, etc.) but this would require additional experiments for performance evaluation.

\smallskip\noindent\textbf{Future work.} 
Our structured queries (keywords, templates, prototypes) are written in English and then used to rank informative messages in other languages; it is possible that having multilingual queries could improve the performance of the system. 
We are also summarizing information from social media and comparing it against official reports while using both sources (e.g., performing contrastive summarization) could help to generate summaries with extended informativeness and improved veracity. 
Our method assumes that messages are geolocated through some external method, while we could make our method aware of the location of the messages and/or incorporate location elements into the queries.
An envisioned next step for our research is to aggregate and consolidate extracted claims belonging to the same geographical region or event and present these summaries through a map and other information products, potentially integrating them with other information sources.

\smallskip\noindent\textbf{Reproducibility.} 
All of the data and code used for the experiments presented in this paper will be made freely available in a public repository with the camera-ready version of this paper. 
Code and data are available in a repository https://github.com/vitiugin/CLiQS-CM.

\smallskip\noindent\textbf{Acknowledgments.} 
This work has been partially supported by:
"la Caixa" Foundation (ID 100010434), under the agreement LCF/PR/PR16/51110009;
the Ministry of Science and Innovation of Spain with project "COMCRISIS", reference code PID2019-109064GB-I00;
and
the EU-funded "SoBigData++" project, under Grant Agreement 871042.

%Bibliography
\bibliographystyle{unsrt}  
\bibliography{main}

\end{document}